# Emission Variation of a Long-period Pulsar Discovered by the Five-hundred-meter Aperture Spherical Radio Telescope (FAST)

H. M. Tedila[1,2,3], R. Yuen[1,4], N. Wang[1,4,5], J. P. Yuan[1,4,5], Z. G. Wen[1,4,5], W. M. Yan[1,4,5], S. Q. Wang[1,2], S. J. Dang[6,7], D. Li[2,8,9], P. Wang[8], W. W. Zhu[8], J. R. Niu[2,8], C. C. Miao[2,8], M. Y. Xue[8], L. Zhang[2,10,11], Z. Y. Tu[1,2], R. Rejep[1,2], and J. T. Xie[1,2]
FAST Collaboration
[1] Xinjiang Astronomical Observatory, Chinese Academy of Sciences, 150 Science 1-Street, Urumqi, Xinjiang 830011, People's Republic of China; ryuen@xao.ac.cn, na.wang@xao.ac.cn
[2] University of Chinese Academy of Sciences, 19A Yuquan Road, 100049 Beijing, People's Republic of China
[3] Arba Minch University, Arba Minch 21, Ethiopia
[4] Key Laboratory of Radio Astronomy, Chinese Academy of Sciences, Nanjing, 210008, People's Republic of China
[5] Xinjiang Key Laboratory of Radio Astrophysics, 150 Science1-Street, Urumqi, Xinjiang, 830011, People's Republic of China
[6] School of Physics and Electronic Science, Guizhou Normal University, Guiyang 550001, People's Republic of China
[7] Guizhou Provincial Key Laboratory of Radio Astronomy and Data Processing, Guizhou Normal University, Guiyang 550001, People's Republic of China
[8] CAS Key Laboratory of FAST, NAOC, Chinese Academy of Sciences, Beijing 100101, People's Republic of China
[9] NAOC-UKZN Computational Astrophysics Centre, University of KwaZulu-Natal, Durban 4000, South Africa
[10] National Astronomical Observatories, Chinese Academy of Sciences, A20 Datun Road, Chaoyang District, Beijing 100101, People's Republic of China
[11] CSIRO Astronomy and Space Science, PO Box 76, Epping, NSW 1710, Australia
Received 2021 November 25; revised 2022 March 18; accepted 2022 March 19; published 2022 April 26

## Abstract

We report on the variation in the single-pulse emission from PSR J1900+4221 (CRAFTS 19C10) observed at frequency centered at 1.25 GHz using the Five-hundred-meter Aperture Spherical radio Telescope. The integrated pulse profile shows two distinct components, referred to here as the leading and trailing components, with the latter component also containing a third weak component. The single-pulse sequence reveals different emissions demonstrating as nulling, regular, and bright pulses, each with a particular abundance and duration distribution. There also exists pulses that follow a log-normal distribution suggesting the possibility of another emission, in which the pulsar is radiating weakly. Changes in the profile shape are seen across different emissions. We examine the emission variations in the leading and trailing components collectively and separately, and find moderate correlation between the two components. The inclination angle is estimated to be about $7°$ based on pulse-width, and we discuss that nulling in this pulsar does not seem to show correlation with age and rotation period.

*Unified Astronomy Thesaurus concepts:* Pulsars (1306)

## 1. Introduction

Radio pulsars are well known for exceptionally regular emission due to the stability in their rotation. However, at the individual pulses, large variations in pulse shape and flux are detected (Penny 1982; Patt et al. 1999), and the pulse energy fluctuates from pulse to pulse in all radio pulsars (Manchester & Taylor 1977; Edwards & Stappers 2003). When a sequence of pulses is displayed, the fluctuation reveals as different emission phenomena, which may appear quasi-periodic and affecting either whole or part of the on-pulse window (Backer 1970; Rankin & Wright 2008). The emission phenomena may include profile mode-changing (e.g., Bartel & Hankins 1982; Wang et al. 2007), nulling (e.g., Backer 1970; Rankin 1986; Wang et al. 2007), subpulse drifting (Drake & Craft 1968; Edwards & Stappers 2003; Weltevrede et al. 2006), and giant pulses (e.g., Staelin & Reifenstein 1968; Hankins et al. 2003; Kuzmin & Ershov 2006; Sun et al. 2021).

Quite commonly, radio pulsars cease to emit radio waves for several pulse periods or minutes (Backer 1970; Rankin 1986; Wang et al. 2007; Young et al. 2015). Known as nulling, the phenomenon may be considered as an extreme form of profile mode-changing (Backer 1970; Wang et al. 2007; Zhang et al. 2019). The latter phenomenon exhibits as a sudden change in the stable pulse profile from one shape to another at an observing frequency (Lyne et al. 1971; Manchester & Taylor 1977; Bartel & Hankins 1982). Ritchings (1976) suggested that the fraction of pulses without detectable emission (nulling fraction, hereafter NF) was correlated with the age of the pulsar, and Biggs (1992) proposed that the NF was associated with the pulsar rotation period.

For a significant proportion of the observed pulsars (Weltevrede et al. 2006, 2007), a correlation in the change of the subpulse longitude in one pulse with the previous and subsequent pulses (Drake & Craft 1968; Edwards & Stappers 2003; Weltevrede et al. 2006) is also detected. Known as drifting subpulses, the phenomenon demonstrates as a systematic flow of subpulses across the pulse window resulting in a drift pattern in successive pulses (Manchester & Taylor 1977). The diversity in the observed emission phenomena indicates variations in the properties of the emission region and the differences in the sight-line path that traverses it in different pulsars. In addition, global magentospheric configuration may also play a role, as with subpulse drifting, which is believed to be set by the accelerating potential in the polar gap.

For some pulsars that exhibit multiple single-pulse phenomena, correlations between the different phenomena are usually observed (van Leeuwen et al. 2003; Gajjar et al. 2014). An example relates to PSR B0943+10, which shows correlation between subpulse drifting properties and the shape of the integrated pulse profile observed at different frequencies (Suleymanova & Rankin 2009). The observation of PSR B0809+74 reveals that the subpulse drift rate changes when the pulsar nulls with an accompanying change in the phase of the post-null average profile (van Leeuwen et al. 2003). In addition, a variation in single pulses can produce longer lasting effects to the observed emission. For example, the changes in emission due to nulling and subpulse drifting will also cause the shape of the integrated pulse profile to change (Redman et al. 2005; Gajjar et al. 2017).

In this paper, we report the emission properties of PSR J1900+4221 based on a single-pulse observation of a two-hour duration using the Five-hundred-meter Aperture Spherical radio Telescope (FAST) in Guizhou, China. FAST is a single-dish telescope possessing three times higher sensitivity and a factor of ten times greater surveying speed than the Arecibo telescope (Nan & Li 2013). The pulsar was discovered by FAST on 2018 August 2 in the drift-scan observation using the 19 beam receiver. It has a rotation period of 4.34 s, and dispersion measure of $72 \, \text{cm}^{-3} \, \text{pc}$. The pulsar was confirmed by the follow-up observation conducted on 2019 January 22,[12] and has no previous publication for its emission properties.

The paper is organized as follows. Our observation and data processing are presented in Section 2, and the results are described in Section 3. Analysis of the emission from the pulsar is presented in Sections 4 and 5. We discuss and summarize our results in Section 6.

## 2. Observation and Data Processing

The single-pulse observation of PSR J1900+4221 was performed for two hours on 2020 October 13 with FAST using the 19 beam receiver and the Reconfigurable Open Architecture Computing Hardware–version2 (ROACH2) signal processor (Jiang et al. 2019). More information on the receiver and the backend systems are given by Jiang et al. (2020). The observation was performed over the frequency range from 1.05 to 1.45 GHz, with time and frequency resolutions of 49.152 $\mu$s and 0.488 MHz, respectively. The data was then recorded with 1024 frequency channels in search-mode PSRFITS format (Hotan et al. 2004).

The ephemerides for newly discovered pulsars are usually not precise. For PSR J1900+4221, we used TEMPO2 (Hobbs et al. 2006) software package to obtain the best-fit ephemeris to correct for the rotation period. The DSPSR software package (van Straten & Bailes 2011) was then used to dedisperse and produce the single-pulse integration. Spectral band edges were zapped, and the interference was removed from the data using paz and pazi, respectively, in the PSRCHIVE software package (Hotan et al. 2004). The data were then scrunched in frequency and polarization. Next, the observed pulses were calibrated based on the off-pulse noise using the radiometer equation (Dicke 1982; Wen et al. 2021),

$$S_{\text{av}} = \frac{S_{\text{sys}}}{\sqrt{n_{\text{p}} f_{\text{b}} t_{\text{s}}}}. \quad (1)$$

Here, $n_{\text{p}} = 4$, $f_{\text{b}} = 400 \, \text{MHz}$ and $t_{\text{s}} = 49.152 \, \mu\text{s}$ are the number of polarization, effective frequency bandwidth, and sampling time, respectively, and $S_{\text{sys}} \approx 1 \, \text{Jy}$ is the system equivalent flux density of FAST (Yu et al. 2017; Wang et al. 2021). Flux calibration was then performed on all single pulses by multiplying with $S_{\text{av}}$. We obtained 1628 single pulses.

## 3. Single-pulse Emission

Figure 1 shows the single-pulse sequence from the whole observation in the upper panel, and the corresponding integrated pulse profile is presented in the lower panel. The integrated pulse profile is composed of two main components, referred to here as the leading and trailing components, which are signified as I and III, respectively. Further examination reveals that the trailing component is also composed of another weaker component, whose peak is designated by II and located about 2° earlier in longitudinal phase from the peak of component III. The greatest peak intensity is found in component I.

It is apparent from the single-pulse sequence that emission from this pulsar exhibits repeating patterns of regular and null pulses. There are also occasional pulses whose intensities are much brighter than those of the ordinary pulses. Two examples are seen at pulse numbers around 350 and 1050 in the trailing and leading components, respectively.

Figure 2 shows the emission variation of the pulsar in three different pulse subsequences. Examples are shown in blue, yellow, and green rectangles in the middle and the rightmost subsequences, where the emission from the leading and trailing components appear *synchronized* in the sense that variation in the pulse intensity appears in step between the two components. This emission pattern lasts for almost 100 pulse periods in both subsequences. Another emission pattern is shown in the leftmost plot between pulse numbers around 400–420 enclosed by the rectangle in red. The emission appears *nonsynchronized* in the sense that the pulse intensity in successive pulses from the two components is noticeably different and dominated by only one component, which is the leading component in this case. Variations can also occur within the synchronized pattern. This is shown in the rightmost plot for pulse numbers around 810–835, enclosed by the yellow rectangle, where emission is detected from both components in blocks of four to ten consecutive pulse periods separated by two episodes of weak emission. Then, the duration reduces to about 2–4 consecutive pulse periods for the next 50 pulses between pulse numbers 840–890 in the green rectangle. In addition, there also exists pulses that appear to cover a broader longitudinal phase and emit weakly. This is seen with some of the single pulses in the rightmost subsequence enclosed by the rectangle in green between pulse numbers 850 and 890. Occasionally, pulse emission with significantly greater peak intensity than average is detected, as shown in Figure 1. We identify the pulses as bright pulses, with each possessing a peak intensity that exceeds the mean intensity by a factor of ten.

We determine the pulse energy distribution by following the procedure outlined by Ritchings (1976). In this approach, the on-pulse energy of a pulse is determined by integrating the intensity within the on-pulse window subtracting the corresponding off-pulse baseline noise. The off-pulse energy of a pulse is calculated from an off-pulse region with a width that is the same as the on-pulse window. The mean on-pulse energy is then used to normalize all the on- and off-pulse energies. Figure 3 presents the histograms of pulse energy distributions in log scale for the off-pulse window (red) and the on-pulse window (black). As shown in the inset, the on-pulse energy distribution can be fitted by two Gaussian components with one peak at zero (blue), representing the Gaussian random noise from the system, and the other at higher energy (green). A log-normal distribution (yellow) is used to fit the data at low energy, which indicates that a possible different emission mechanism is involved. The equations for the Gaussian and log-normal distributions are

$$F(E) = \frac{\alpha}{\sigma\sqrt{2\pi}} \exp\left(-\frac{(E-\mu)^2}{2\sigma^2}\right) \quad (2)$$

and

$$F(E) = \frac{\alpha}{E\sigma\sqrt{2\pi}} \exp\left(-\frac{(\ln(E)-\mu)^2}{2\sigma^2}\right), \quad (3)$$

respectively, where $\alpha$ is the shape parameter, $\mu$ is the mean, and $\sigma$ represents the standard deviation of the distribution. Table 1 lists the parameters used to fit each distribution.

---

[12] PSR J1900+4221 was discovered during the pilot scans of the Commensal Radio Astronomy FAST Survey (CRAFTS; Li et al. 2018). It was confirmed as the 10th discovery from the 19 beam portion of the survey and is listed as CRAFTS 19C10 on the survey webpage 1. https://crafts.bao.ac.cn/pulsar/.

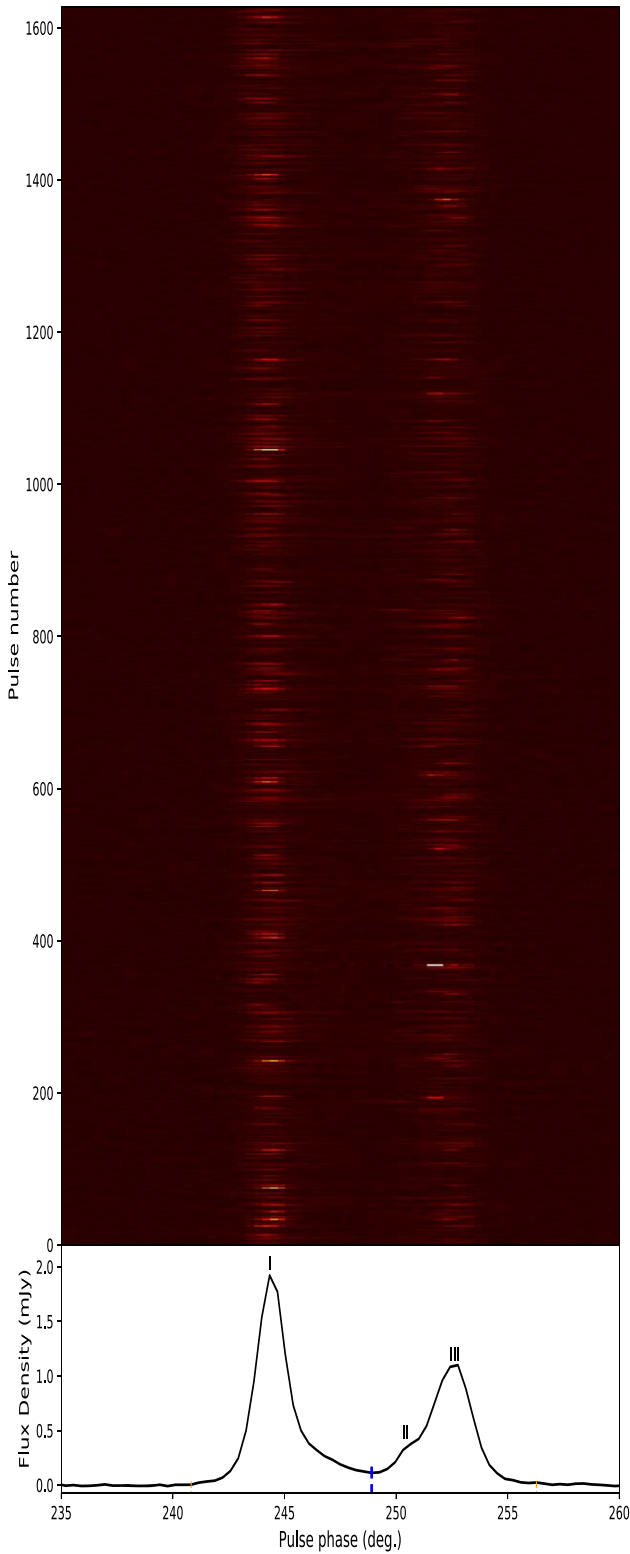

**Figure 1.** The sequence of 1628 single pulses from our observation is presented in the upper panel, and the integrated pulse profile is shown in the lower panel. The blue vertical dashed line in the lower panel indicates the point of lowest intensity in the profile between the leading and trailing components, which is used to separate the two components.

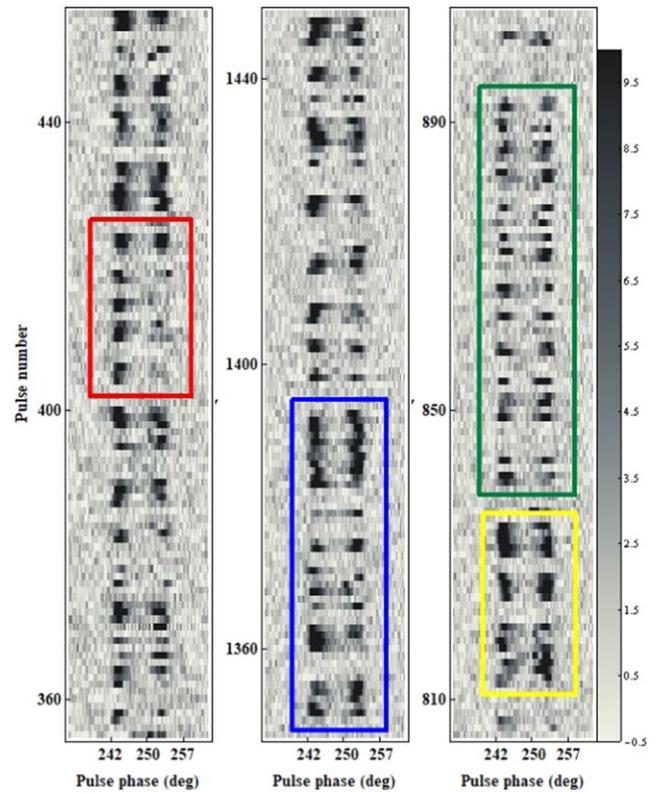

**Figure 2.** Three subsequences of pulses showing the emission variation in the leading and trailing components at different times. Pulses in the red rectangle represent *nonsynchronized* emission (see main text), and pulses in the yellow and green rectangles corresponding to *synchronized* emission of long and short durations, respectively. The blue rectangle contains pulses of *synchronized* emission of different durations.

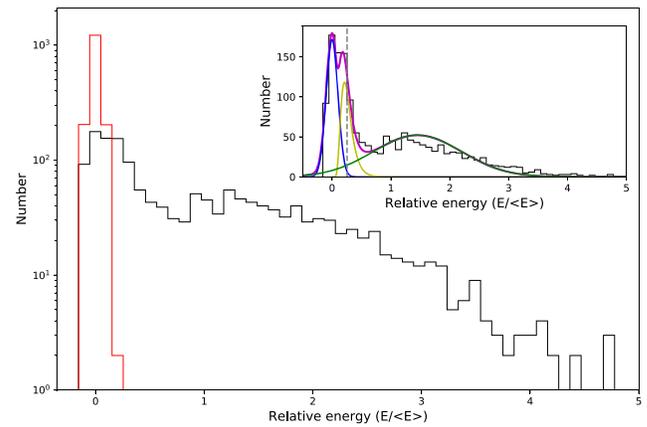

**Figure 3.** Pulse energy distributions for the on-pulse (black histogram) and off-pulse (red histogram) regions in log scale. The original (not log scale) on-pulse distribution is shown in the inset. Here, the solid line in magenta represents the fitting for the on-pulse energy distribution based on combining the two Gaussian components (blue and green) and the log-normal distribution (yellow). The vertical gray dashed line represents the threshold energy value for Gaussian fitting to both null pulses and off-pulse regions. The x-axis is normalized by the mean on-pulse energy.

**Table 1**
List of the Parameters Used to Fit the Energy Distributions for the Null and Regular Pulses (Gaussian), and for the Weak Pulses (Log-normal) in Figure 3

| Parameters | Null | Weak | Regular |
|---|---|---|---|
| $\alpha$ | 43 | 27 | 98 |
| $\mu$ | 0 | −1.4 | 1.5 |
| $\sigma$ | 0.1 | 0.4 | 0.75 |

### 4. Identification of the Different Emissions

To distinguish the different emissions in the pulsar, we employ the method suggested by Bhattacharyya et al. (2010) and Yan et al. (2019, 2020). The approach involves determining the uncertainty threshold in the on-pulse energy for each single pulse using $\eta_{on} = \sqrt{n_{on}}\,\sigma_{off}$. Here, $n_{on}$ is the bin number for the on-pulse region, and $\sigma_{off}$ is the root mean square for the off-pulse region obtained based on the same $n_{on}$. The on-pulse range is determined at about 10% of the maximum intensity of the integrated pulse profile at the beginning and end of the leading and trailing components, respectively. We refer to the pulses with pulse energy less than or equal to $1 \times \eta_{on}$ as null pulses, and the pulses with energy lies between $\eta_{on}$ and $5 \times \eta_{on}$ as weak pulses. The regular pulses possess pulse energy that is greater than or equal to $5 \times \eta_{on}$.

#### 4.1. Emission Variations

Figure 4 shows the integrated pulse profiles for the regular, weak, and null pulses all normalized to that from the whole observation. The profile for regular pulses largely resembles that from the whole observation. However, the profile integrated from the weak pulses is different. It exhibits similar peak intensity for components I, II, and III, and the profile width at 10% intensity level is wider than that from the regular pulses by nearly 40%. The upper panel in Table 2 gives the statistical information for the occurrences of the three different pulse emissions. The regular pulses are the most dominant emission, at about 55%, as compared to the null and weak pulses combined. However, the occurrence of weak pulses is slightly more than that for the null pulses.

The total durations for the null, weak, and regular pulses are different, with the regular pulses dominating, as shown in the last column in Table 2. Figure 5 shows the histograms of the duration distribution for the null, weak, and regular pulses.

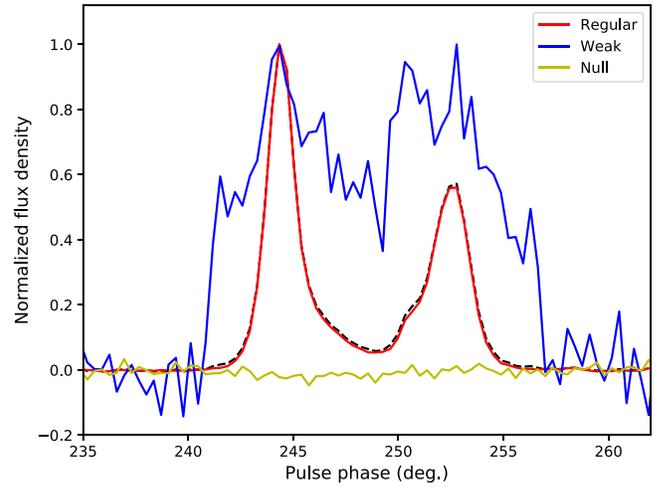

**Figure 4.** Plot showing the integrated pulse profiles for the regular (red), weak (blue), and null (yellow) pulses, with the first two normalized to the integrated pulse profile obtained from the entire observation (dotted-black).

**Table 2**
Occurrences for the Null (N), Weak (W), and Regular (R) Pulses in Different Components from Our Observation

| Component | Emission | Pulses | Abundance (%) | Duration |
|---|---|---|---|---|
| Both | N | 342 | 21.0 | 24.7 |
|  | W | 391 | 24.0 | 28.3 |
|  | R | 895 | 55.0 | 64.7 |
| Leading | N | 422 | 25.9 | 30.5 |
|  | W | 421 | 25.9 | 30.5 |
|  | R | 785 | 48.2 | 56.8 |
| Trailing | N | 414 | 25.4 | 29.9 |
|  | W | 466 | 28.6 | 33.7 |
|  | R | 748 | 46.0 | 54.1 |

**Note.** The total duration is expressed in minutes.

They indicate that the null and weak pulses are mostly single-pulse events. However, the maximum duration of regular pulses can reach up to eleven pulse periods, with a mean value, $m_R$, of three pulse periods.

#### 4.2. Pulse Emission around Nulls

There are 249 null blocks in total of 342 null pulses in the two-hour observation. We determine the fraction of null pulses, or the NF, for the pulsar based on the histograms in Figure 3. Following the method introduced by Ritchings (1976) and Wang et al. (2007), an increasing fraction of noise probability distribution in the off-pulse histogram centered on zero energy was subtracted from the distribution of the observed pulse in the on-pulse histogram until the sum of the difference counts in bins with E < 0 was zero. The uncertainty in the NF is determined by $\sqrt{n_p}/N$, where $n_p$ is the number of null pulses and $N$ is the total number of observed pulses (Wang et al. 2007; Wen et al. 2016). We obtain NF = 22.2% ± 1.1% for this pulsar. Table 3 shows the single-pulse emission immediately before and after each null. We find that regular pulses occur consistently more than the weak pulses in both cases, and their occurrences do not change significantly before and after the nulls. We also detected two instances of bright pulses (see Section 5.2) before the nulls, and one instance after a null.

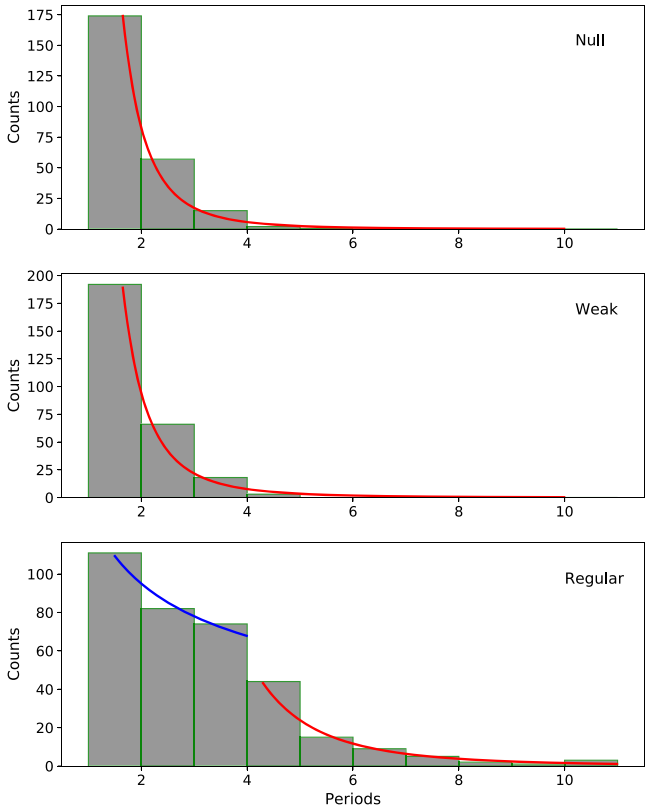

**Figure 5.** Histograms for the durations of the null (upper), weak (middle), and regular (lower) pulses from both components. The red curve in each of the upper and middle plots represents the best-fit power law to each distribution, whereas the distribution in the lower plot requires two power laws (red and blue) to fit.

**Table 3**
Statistics for the Single-pulse Emission before and after the Nulls

| Emission | Before | After |
| --- | --- | --- |
| Regular (R) | 159 | 154 |
| Weak (W) | 88 | 94 |
| Bright | 2 | 1 |

Figure 6 shows the integrated pulse profiles obtained from all the single pulses immediately before and after all the null blocks. The profile peaks of the leading and trailing components before the nulls are both higher than those integrated from pulses after nulls.

## 5. Emission Variations in the Leading and Trailing Components

Different emission properties can be identified when considering the leading and trailing components separately. In this section, we apply the same criteria for identification of the different emission, as stated in Section 4, to the leading (I) and the trailing (II & III) components separately. The pulse energy represents that within the respective component range separated by the blue vertical dashed line as indicated in Figure 1.

As shown in the middle and lower panels in Table 2, the regular pulses remain the most dominant emission but at slightly different occurrence rates with more regular pulses detected in the leading component. In addition, roughly equal number of null and weak pulses are observed in the leading component, whereas weak pulses are slightly more than the null pulses in the trailing component. The total duration for the null, weak, and regular pulses is different for the leading and trailing components. Figure 7 shows the histograms of the duration distributions for the three pulses in the leading and trailing components. They indicate that the null and weak pulses occur mostly as single-pulse events. However, null pulses can last up to a maximum of five consecutive pulse periods in the leading component, and the duration for weak pulses varies up to six pulse periods in the trailing component. The duration for the regular pulses can last up to ten pulse periods with a large proportion exhibiting as single-pulse event in both components.

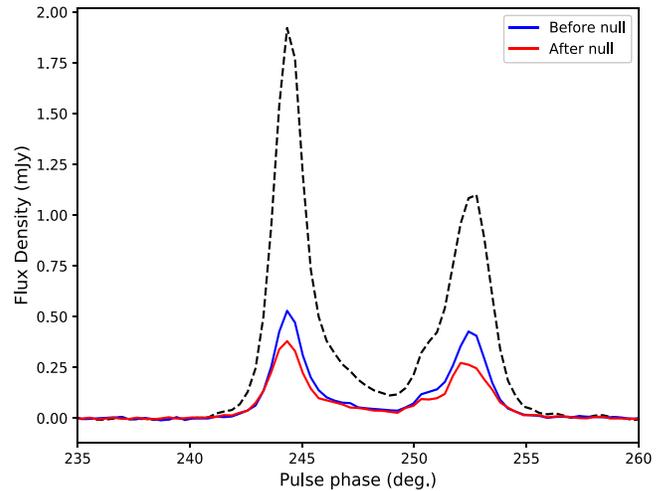

**Figure 6.** Plot showing the integrated pulse profiles obtained from the one pulse before (blue) and after (red) each null block against that from the whole observation (dashed).

### 5.1. Regular Pulse Emission

Figure 2 shows that emission from the regular pulses exhibits two different patterns demonstrated as different lengths of consecutive emission. The duration of consecutive pulse emission in one pattern appears longer, as indicated by the yellow rectangle, than the other pattern that shows a shorter duration, as shown by the green rectangle. From the duration distribution for the regular pulses shown in Figure 5, a noticeable step change between three and four is observed (around $m_R$) requiring two power laws to fit the distribution, as opposed to the distributions for null and weak pulses each involving only a single power law for fitting. The power laws that fit the duration distributions for the null pulses and weak pulses have exponents of $-3.87$ and $-3.63$, respectively. The two power laws used to fit the duration distribution for the regular pulses have exponents of $-0.48$ (blue) and $-3.93$ (red).

We separate the regular pulses into two groups, named $G$ and $L$, based on the consecutive emission being $\geqslant m_R$ or $< m_R$, respectively. Table 4 lists the statistics for their occurrences. From the 785 regular pulses in the leading component, 469 pulses belong to group $G$ indicating that a higher proportion of regular pulses was found with consecutive pulse number $\geqslant m_R$. The similar relation was also identified in the trailing component. The overall similarity in statistics for each group between the two components suggests that the emission is correlated between the two components. Next, we identify only

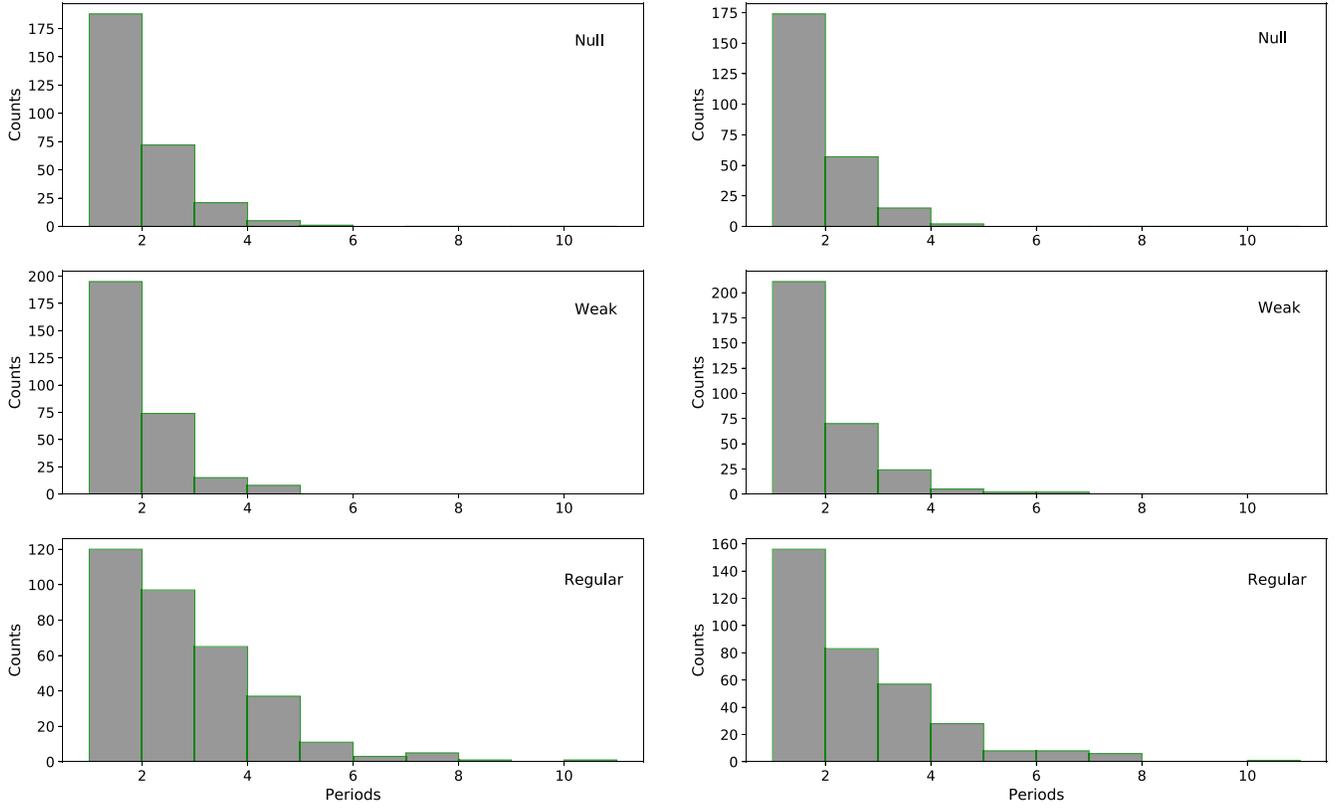

**Figure 7.** Similar to Figure 5 but for the leading (left) and trailing (right) components.

**Table 4**
Occurrences of Regular Pulses in the Leading and Trailing Components

| Component | Regular Pulses | Pattern | Pulses | Abundance (%) |
|---|---|---|---|---|
| Leading | 785 | G | 469 | 59.8 |
|  |  | L | 316 | 40.2 |
| Trailing | 748 | G | 424 | 56.7 |
|  |  | L | 324 | 43.3 |

**Note.** The emission patterns labeled with $G$ and $L$ signify regular pulses that occur for consecutive pulse number $\geqslant m_R$ and $< m_R$, respectively.

the consecutive and synchronized pulses in the two groups from both components and perform pulse integration, and the results are shown in Figure 8. While the shape of the profile for group $G$ resembles that of the whole observation, the component II is more noticeable in the profile from group $L$. In addition, the profile flux density for group $G$ is larger than that for group $L$. The ratios of the peak flux density between components I and III for profiles in group $G$ and $L$ are 1.87 and 1.55, respectively. We verify the former ratio based on different integrated pulse profiles obtained using similar number of pulses as in group $L$. They all give ratios that are consistently greater than 1.8. This indicates that the emission properties are different between the two groups.

### 5.2. Bright Pulses

A total of 12 instances of bright pulses, or about 0.7% of all the pulses, were identified in the two-hour observation. Two were detected in the trailing component, and the remaining are all found in the leading component. Bright pulses are single-pulse events. Figure 9 shows examples of the bright pulses detected at four different times. For (a), (b), and (c), the bright pulses are emitted from the leading component, whereas the bright pulse in (d) comes from the trailing component.

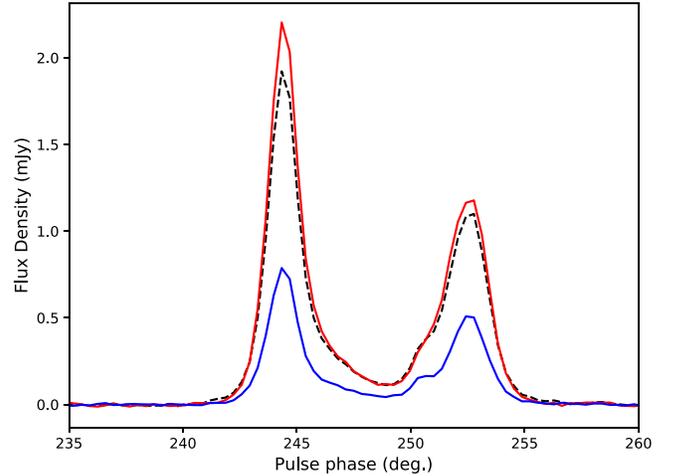

**Figure 8.** Plot showing the integrated pulse profiles for the regular pulses with consecutive pulse duration $< m_R$ (blue) and $\geqslant m_R$ (red), respectively, from both components relative to the integrated pulse profile of the entire observation data (dashed black).

From our observation, a bright pulse can emit from either the leading or the trailing component as shown in Figure 9. During the event, the component that is not emitting a bright pulse may have a peak intensity that is greater than (as in (a)) or roughly equal to (as in (c) and (d)) that of the integrated pulse profile. However, emission is not detectable from the trailing component during a bright-pulse event as shown in (b). The bright pulse with the highest peak intensity occurred in the trailing component from our observation. Figure 10 shows the integrated pulse profiles for the bright pulses found in the leading and trailing components, and the locations of the corresponding peak intensity relative to the integrated pulse profile from the whole observation. For bright pulses in the leading component, the peaks of the intensity occur at the pulse phases that mostly align with the peak of component I. However, the peak intensity of the bright pulses in the trailing component are found located slightly at earlier longitudinal phases between components II and III.

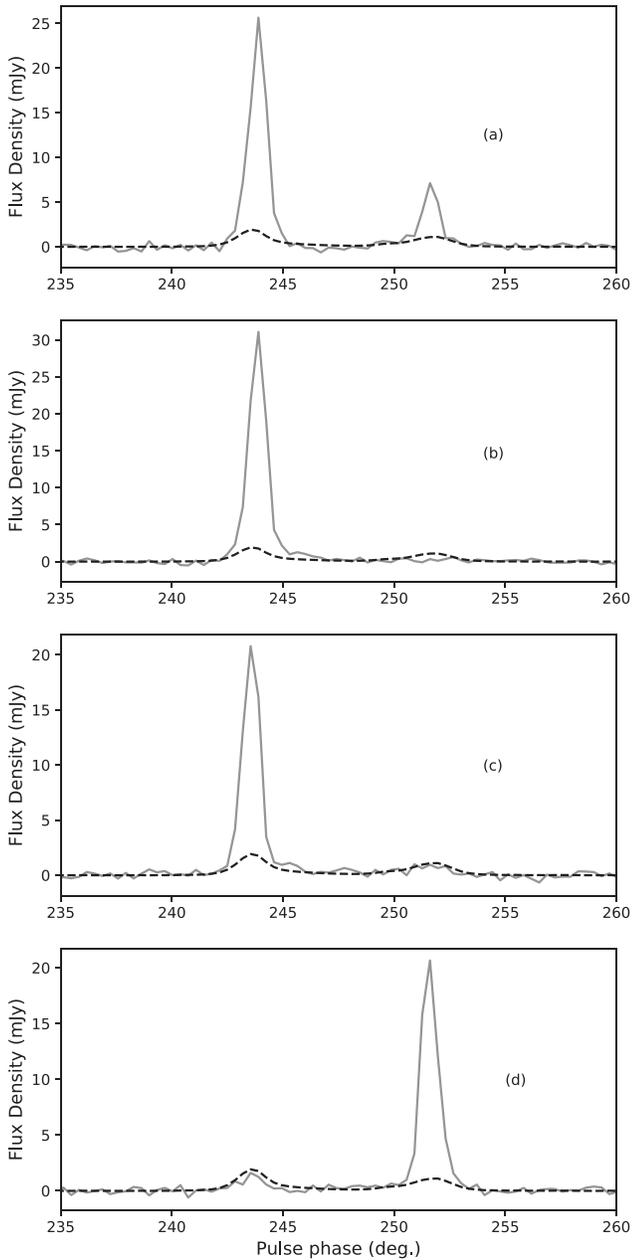

**Figure 9.** Four examples of bright pulse (solid gray) detected at pulse numbers 35 (a), 76 (b), 610 (c), and 1374 (d). The dashed line in each plot indicates the integrated pulse profile from the whole observation.

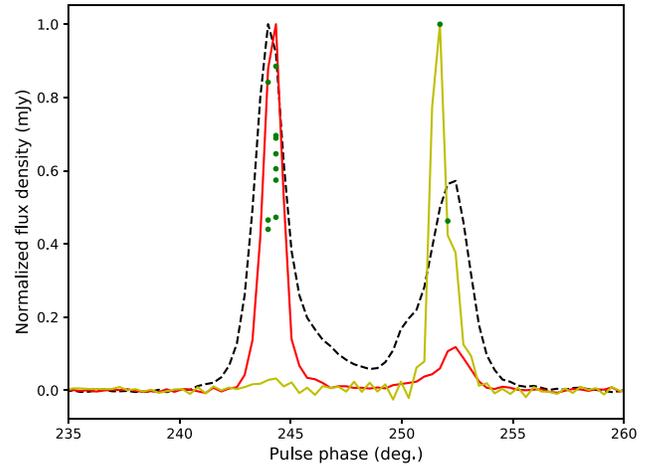

**Figure 10.** Plot showing the integrated pulse profiles for the bright pulses found in the leading (red) and trailing (yellow) components, and the locations for the peak intensity (green dots) relative to the integrated pulse profile (dashed black).

### 5.3. Apparent Movement of Subpulses

We identify several instances where the subpulses in the leading and trailing components exhibit systematic changes in the longitudinal phase in consecutive pulses producing an effect of what appears to be movement across the respective component window. To avoid misidentification with the intrinsic emission variation in individual pulses, only the movements that last for three pulse periods or more are considered. The phenomenon is more commonly seen in the trailing component. The result of change exhibits as subpulses moving in either positive or negative direction corresponding to the apparent movement of subpulses to later or earlier longitudinal phases, respectively. In addition, the phenomenon can occur concurrently in both components and in opposite directions.[13] Figure 11 shows four instances identified in the observation. The longest duration for the movement is six pulse periods, and the average duration is about three pulse periods for both components. The phenomenon occurs mainly in regular pulses, and bright pulses are not detected in any of the events. In addition, the single pulse before or after the subpulse movement is either a null or a weak pulse. An example of null pulses detected before and after the event is between pulse numbers 823 and 826 as shown in the second plot (from top) in Figure 11. In our observation, the movement of subpulses are not preceded or followed by a regular or bright pulse.

### 5.4. Correlation between the Two Components

The analyses in this section show that emission from the leading and trailing components are occasionally different. The correlation in the emission variations between the leading and trailing components is demonstrated in Figure 12, which shows the cross-correlation in pulse energy between the two components. The result reveals a maximum correlation coefficient ($r$) of 0.48 at zero lag, indicating a moderate linear relationship between the two components. In addition, the rapid decrease in the value of $r$ on both sides suggests the lack of periodicity for the emission variation in both components.

---

[13] We note here that the movements of subpulse components seen here are more diverse than those of another CRAFTS pulsar, namely PSR J1926-0652 (Zhang et al. 2019).

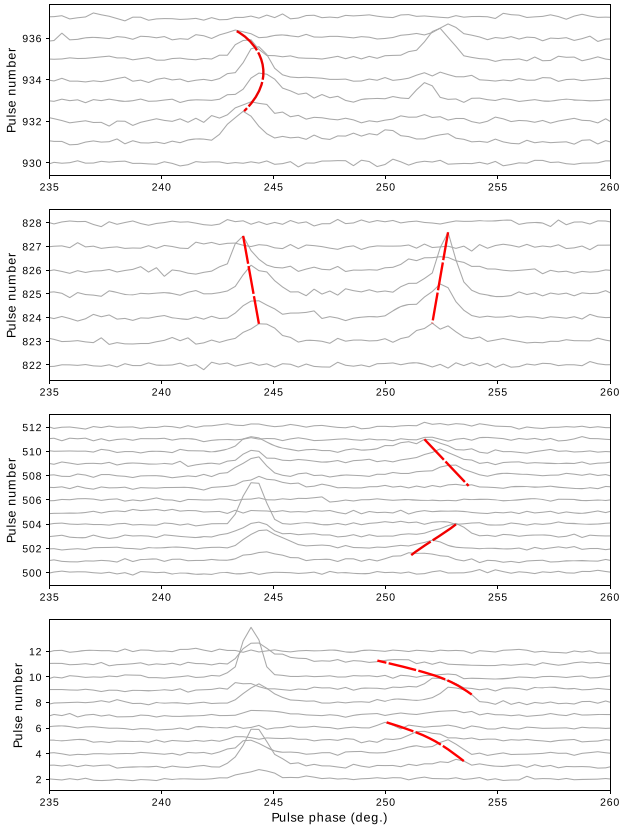

**Figure 11.** Four different single-pulse sequences showing the apparent movement of subpulses across the pulse window of the respective component.

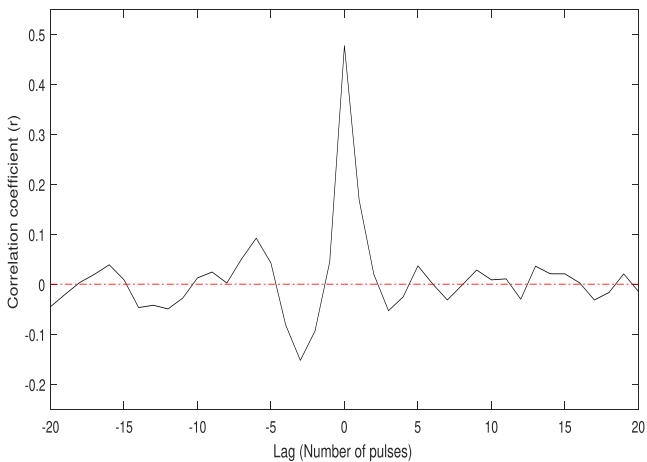

**Figure 12.** The cross-correlation in the pulse energy between the leading and trailing components.

### 6. Summary and Discussion

We have reported on the emission variation in PSR J1900 +4221 for the first time based on a two-hour single-pulse observation that was performed on 2020 October 13 using FAST at a frequency centered at 1.25 GHz. The pulsar has a relatively long rotation period of 4.34 s, and exhibits variation in the single-pulse emission. The latter demonstrates as nulling, weak, and regular pulses, occasional bright pulses and apparent movement of subpulses. We find that the regular pulses exhibit two different emission patterns, each with a different pulse profile showing that the emission properties are different. We obtain a NF of 22.2% ± 1.1% for this pulsar.

In general, there are two proposals for the cause of nulling. One model relates the event to changes in the viewing geometry such that the line of sight no longer cuts the emission beam (Timokhin 2010). The suggested change is in the rearrangement of the open field lines causing the emission beam to reorient. The other model involves temporary loss of the coherent radio emission in relation to the plasma production in the polar gap (Filippenko & Radhakrishnan 1982). In our observation, the emission from the single pulse immediately before and after a null shows preference for relatively high pulse energy of a regular pulse or even a bright pulse. From the similar duration for nulls between the leading and trailing components, as shown in Table 2 and Figure 7, the process for the cessation of emission appears concurrent across the observable emission region. This indicates that the stopping and restarting of the plasma generation, or the changing and restoring of the magnetic field structure, both take place across a significant size of the emission region within a timescale of a few seconds.

NF was proposed in positive correlation with the age (Ritchings 1976) and the rotation period (Biggs 1992) of a pulsar as mentioned in the introduction section. The evolution of the pulsar age is suggested in correlation with the inclination angle, $\alpha$. From consideration of energy loss through magnetic dipole radiation (Manchester & Taylor 1977; Lyubarskii & Kirk 2001), the evolution of $\alpha$ is from large to small. However, consideration of the longitudinal current flow and the pair production in the magnetosphere (Beskin et al. 1988) suggests that the change of $\alpha$ is from small to large as a pulsar ages. Furthermore, pulsar age is also related to the rotation period of a pulsar. In the Ruderman & Sutherland (1975) model, the failure to maintain the Goldreich & Julian (1969) charge density ($\rho_{GJ}$) along open field lines implies the development of a potential to accelerate charged particles in the vacuum gap. This accelerating potential reduces as the pulsar rotation slows down, and pulsar evolution is from short to long rotation periods with radio pulsar death being signified by the suppression of the pair production at a large age (Arons 2000; Zhang et al. 2000; Harding & Muslimov 2002). The relation of the rotation period of a pulsar to its age implies that the former is also related to $\alpha$.

The relatively long rotation period of PSR J1900+4221 represents an important case for offering definitive information on the correlation between nulling and $\alpha$. Using the method that based on the pulse-width at about 50% intensity (Rankin 1990; Maciesiak et al. 2011), we obtain $\alpha \approx 7°$ for this pulsar. With the relatively long rotation period and a small inclination angle, the evolution of this pulsar appears to suggest that the energy loss is due to magnetic dipole radiation. However, the estimated 22.2% of NF appears normal among the nulling pulsars (Wang et al. 2007), and so it does not seem to show a strong correlation with either the pulsar age or the rotation period. This also implies that NF is not correlated with $\alpha$ for this pulsar. However, there are different methods for estimating the inclination angle, with some yielding different results. For example, the profile widths of PSRs B0823+26 and B1819-22 are both about 15° at 21 cm observing frequency (Weltevrede et al. 2006), but the predicted inclination angles are vastly different at 98°.9 (Everett & Weisberg 2001) and 17° (Lyne & Manchester 1988), respectively. It may be that our estimation of $\alpha$ is not the true value for PSR J1900+4221. It follows that determining the value of $\alpha$ will provide a specific way to differentiate between different models for nulling in this pulsar, which will be the topic of our future paper.

We summarize the emission features in PSR J1900+4221 below.

1. The pulsar has a rotation period of 4.34 s, and displays an integrated pulse profile with three discernible emission components.
2. Emission variation in the pulsar demonstrates as null, weak, and regular pulses, with the regular pulses (above $5\sigma$ of the noise) possessing the highest occurrence rate.
3. Regular pulses are emitted in two patterns demonstrating as long ($\geqslant 3$ pulse periods) and short ($<3$ pulse periods) consecutive emission, with the average pulse energy being larger in the former.
4. The integrated pulse profile from the weak pulses has width that is about 40% broader than that of the whole observation.
5. The leading and trailing components each demonstrates unique emission properties, and the emissions from the two components exhibit moderate correlation.
6. Pulses with peak intensity above ten times of the average are detected. These bright pulses emit from either the leading or the trailing component, with detection being more common in the former component.

This work is partially supported by National Natural Science Foundation of China (NSFC grant Nos. 11988101, U1838109, 11873080, 557 12041301). We thank the XAO pulsar group for valuable discussions. We also thank all the members of the FAST telescope collaboration for the establishment of the projects (project number: ZD2020_06/PT2020_0052), which led to the observations being available. We are grateful to the referee for valuable comments that have improved the presentation of this paper. H.M.T. acknowledges Arba Minch University, and the University of Chinese Academy of Science and CAS-TWAS President's Fellowship Programme that provides the funding and the PhD Scholarship (No. 2019A8016609001). R.Y. is supported by the 2018 Project of Xinjiang Uygur Autonomous Region of China for Flexibly Fetching in Upscale Talents. Z.G.W. is supported by the 2021 project Xinjiang uygur autonomous region of China for Tianshan elites. W.M.Y. is supported by the Key Laboratory of Xinjiang Uygur Autonomous Region No. 2020D04049, the National SKA Program of China No. 2020SKA0120200, the West Light Foundation of Chinese Academy of Sciences (WLFC 2021-XBQNXZ-027), the National Natural Science Foundation of China (NSFC) project (No. 12041303, 12041304) and the CAS Jianzhihua project. D.L. is supported by the 2020 project of Xinjiang uygur autonomous region of China for flexibly fetching in upscale talents.

This work made use of the data from the Five-hundred-meter Aperture Spherical radio Telescope (FAST; Nan et al. 2011). FAST is a Chinese national mega-science facility operated by National Astronomical Observatories, Chinese Academy of Sciences.

*Software:* DSPSR (van Straten & Bailes 2011), PSRCHIVE (Hotan et al. 2004), TEMPO2 (Hobbs et al. 2006).

ORCID iDs

H. M. Tedila ● https://orcid.org/0000-0002-5815-6548
N. Wang ● https://orcid.org/0000-0002-9786-8548
Z. G. Wen ● https://orcid.org/0000-0003-2991-7421
W. M. Yan ● https://orcid.org/0000-0002-7662-3875
S. Q. Wang ● https://orcid.org/0000-0003-4498-6070
D. Li ● https://orcid.org/0000-0003-3010-7661
W. W. Zhu ● https://orcid.org/0000-0001-5105-4058
C. C. Miao ● https://orcid.org/0000-0002-9441-2190
L. Zhang ● https://orcid.org/0000-0001-8539-4237
J. T. Xie ● https://orcid.org/0000-0001-5649-2591